\documentclass[10pt,onecolumn,aps,prd,preprintnumbers,showpacs,superscriptaddress,nofootinbib,amsmath,amssymb,floats,floatfix,showkeys,notitlepage,longbibliography]{revtex4-1}

\usepackage{orcidlink}
\usepackage{comment}
\usepackage{lipsum}
\usepackage{graphicx}
\usepackage{subfigure}
\usepackage{palatino}
\usepackage{sans}
\usepackage{hyperref}
\hypersetup{colorlinks=true,linkcolor=blue,urlcolor=blue,citecolor=blue}
\usepackage[toc,page]{appendix}
\usepackage[normalem]{ulem}
\usepackage{adjustbox}
\usepackage{latexsym}
\usepackage{amsmath}
\usepackage{amssymb}
\usepackage{amsfonts}
\usepackage{dcolumn}
\usepackage{bm}
\usepackage{tikz}
\usetikzlibrary{decorations.pathmorphing}
\usepackage{bigints}
\usepackage{array,tabularx,multirow,booktabs}
\usepackage[tracking=true]{microtype}
\usepackage{soul} %for highlighting
\SetTracking{}{500}
\SetTracking{encoding={*}, shape=sc}{40}
\UseRawInputEncoding %for inputenc error%
\allowdisplaybreaks

\usepackage[utf8]{inputenc}
\usepackage{algorithm}
\usepackage{algorithmicx}
\usepackage{algpseudocode}
\usepackage{xcolor} % For colored text if needed

\begin{document} \sloppy

%\title{Probing the Quantum Vacuum at the Innermost Stable Circular Orbit through the excitation of a Geodesic Atom in a Schwarzschild Spacetime}
\title{The Sound of an Orbit: A Quantum Spectrum at the ISCO}

\author{Reggie C. Pantig \orcidlink{0000-0002-3101-8591}} 
\email{rcpantig@mapua.edu.ph}
\affiliation{Physics Department, School of Foundational Studies and Education, Map\'ua University, 658 Muralla St., Intramuros, Manila 1002, Philippines.}

\author{Ali \"Ovg\"un \orcidlink{0000-0002-9889-342X}}
\email{ali.ovgun@emu.edu.tr}
\affiliation{Physics Department, Eastern Mediterranean University, Famagusta, 99628 North
Cyprus via Mersin 10, Turkiye.}

\begin{abstract}
We investigate the quantum signature of the innermost stable circular orbit (ISCO), a region of profound importance in black hole astrophysics. By modeling an atom as an Unruh-DeWitt detector coupled to a massless scalar field in the Boulware vacuum, we calculate the excitation rate for a detector following a circular geodesic at the ISCO of a Schwarzschild black hole. In stark contrast to the continuous thermal spectra associated with static or infalling observers, our analysis reveals a unique, non-thermal excitation spectrum characterized by a discrete "frequency comb" of sharp, resonant peaks. We show that the locations of these peaks are determined by the orbital frequency at the ISCO, while their intensity increases dramatically as the orbit approaches this final stability boundary. This distinct spectral signature offers a novel theoretical probe of the quantum vacuum in a strong-field gravitational regime and provides a clear distinction between the quantum phenomena experienced by observers on different trajectories. Our findings have potential implications for interpreting the emission spectra from accretion disks and open new avenues for exploring the connection between quantum mechanics and gravity.
\end{abstract}


\maketitle

\section{Introduction} \label{intro}
The synthesis of quantum mechanics and general relativity remains one of the most profound challenges in modern theoretical physics \cite{Rovelli_2004}. In this pursuit, the study of quantum fields in curved spacetimes has emerged as an indispensable tool, offering a window into the fundamental interactions that govern our universe \cite{Birrell:1982ix}. Black holes, as cosmic laboratories of extreme gravity, provide a unique arena to probe the intricate link between quantum phenomena and the fabric of spacetime itself \cite{Bardeen:1973gs}. The behavior of quantum fields in the vicinity of these enigmatic objects not only deepens our understanding of foundational physics but also holds direct relevance for astrophysical observables, such as the celebrated Hawking radiation and the complex dynamics of accretion disks \cite{Page:1974he,Thorne:1974ve,Frank_2002}.

A particularly fascinating frontier in this domain is the interaction between localized quantum systems, such as atoms, and the quantum fields permeating the curved background of a black hole. When an atom is subjected to a strong gravitational field, its internal energy levels are perturbed not only by the spacetime curvature but also by the incessant fluctuations of the quantum vacuum \cite{Audretsch:1995qb}. This interaction can lead to observable phenomena, providing a novel way to probe the structure of both the gravitational field and the quantum vacuum itself. The theoretical underpinnings for such investigations are well-established, rooted in the Unruh effect, which predicts that an accelerated observer in flat spacetime will perceive the vacuum as a thermal bath, and its gravitational analogue, Hawking radiation, where black holes emit thermal particles due to quantum effects near their event horizon \cite{Hawking:1975vcx,Unruh:1976db,Takagi_1986}.

Building on these foundational concepts, previous studies have extensively examined the response of Unruh-DeWitt detectors, which are simplified yet powerful models for atoms, on various trajectories, including those held at static positions or undergoing radial infall toward a black hole. More recently, a quantum optics approach has been successfully applied to investigate the "horizon brightened acceleration radiation" (HBAR) emitted by atoms falling into a black hole, revealing a deep connection between acceleration radiation, thermodynamics, and near-horizon conformal symmetries \cite{Scully:2017utk,Azizi:2021qcu,Azizi:2021yto,Ordonez:2025sqp,Bukhari:2023yuy,Bukhari:2022wyx,Ovgun:2025isv,Pantig:2025okn,Das:2025rzz}.. This framework has proven to be a versatile tool for analyzing such complex quantum interactions.

While these studies have provided invaluable insights, the specific case of a quantum detector following a stable circular geodesic has received comparatively less attention \cite{Garay:1999sk,Louko:2006zv,Casals:2009zh,Hodgkinson:2014iua,Ng:2014kha,Biermann:2020bjh}. This paper seeks to fill this gap by focusing on a region of profound astrophysical and theoretical importance: the ISCO. Located at a radius of ($r = 6M$) for a Schwarzschild black hole, the ISCO represents the final frontier of stable orbital motion, a critical boundary where matter in accretion disks accumulates before its final plunge. Understanding quantum interactions at this precipice is multifaceted \cite{Ottewill:2000qh}. It promises to shed light on how quantum systems behave under extreme gravitational acceleration, potentially revealing novel quantum-gravitational effects. Furthermore, an atom orbiting at the ISCO can act as a sensitive probe of the quantum vacuum structure, offering a way to study effects akin to Hawking radiation without the need for direct observation of the horizon itself \cite{Barbado:2011dx}.

In this work, we compute the excitation probability of a two-level atom moving along a circular geodesic at the ISCO of a Schwarzschild black hole. We model the atom as an Unruh-DeWitt detector coupled to a massless scalar field prepared in the Boulware vacuum state. This choice of vacuum is physically motivated for an observer far from the black hole and allows us to isolate the quantum effects arising purely from the atom's accelerated orbital motion, distinct from the intrinsic thermal radiation of the black hole itself. By evaluating the Wightman function along the atom's unique worldline and applying time-dependent perturbation theory, we determine the transition rate for the atom's excitation. Our central result is the prediction of a discrete, non-thermal emission spectrum, a unique signature of the stable, periodic motion at the ISCO.

The paper is organized as follows. In Sect. \ref{sec2}, we review the theoretical framework, detailing the Schwarzschild geometry, the geodesic equations for the ISCO, and the quantization of a massless scalar field in the Boulware vacuum. In Sect. \ref{sec3}, we present the core of our work: the detailed calculation of the atom's excitation rate, culminating in the derivation of its spectral properties. In Sect. \ref{sec4}, we discuss the physical interpretation of our results, comparing them to other known scenarios and exploring their potential implications. Finally, in Sect. \ref{conc}, we offer our concluding remarks and outline promising avenues for future research.

\section{Theoretical framework} \label{sec2}
To analyze the quantum phenomena experienced by an atom near a black hole, we must first establish the classical spacetime structure and the specific trajectory of the atom. In this section, we outline the Schwarzschild geometry that serves as our gravitational background and derive the worldline for an atom in a circular geodesic at the ISCO.

\subsection{Schwarzschild Spacetime and Geodesics at the ISCO}
The gravitational field of a non-rotating, uncharged, spherically symmetric black hole of mass $M$ is described by the Schwarzschild metric. In geometric units, where the gravitational constant $G$ and the speed of light $c$ are set to unity, the line element in Schwarzschild coordinates $(t, r, \theta, \varphi)$ is given by \cite{Carroll_2019}
\begin{align}
    ds^2 &= -\left(1 - \frac{2M}{r}\right)dt^2 + \left(1 - \frac{2M}{r}\right)^{-1}dr^2 \nonumber \\
    &+ r^2(d\theta^2 + \sin^2\theta d\varphi^2).
\end{align}
Such a metric possesses a coordinate singularity at the event horizon, located at the Schwarzschild radius $r_g = 2M$, which constitutes a causal boundary from which no information can escape to an observer at infinity. To properly analyze wave propagation, particularly in the near-horizon region, it is often advantageous to introduce the tortoise coordinate, $r_*$, defined by the transformation
\begin{equation}
    r_* = r + 2M \ln\left(\frac{r}{2M} - 1\right).
\end{equation}
This coordinate change maps the radial domain outside the horizon, $r \in (2M, \infty)$, to the entire real line, $r_* \in (-\infty, \infty)$, effectively pushing the event horizon to minus infinity. Such a reformulation is particularly useful for casting the radial part of wave equations into a more familiar one-dimensional Schr\"odinger-like form.

Our investigation focuses on the unique dynamics at the ISCO. For a massive test particle, stable circular orbits in the Schwarzschild spacetime can only exist for radii $r > 6M$. The radius $r = 6M$ marks the marginal stability boundary; at this precise location, an orbiting particle can maintain its trajectory, but any infinitesimal inward perturbation will cause it to spiral inevitably into the black hole. This makes the ISCO a region of immense physical interest, representing the inner edge of typical accretion disks around non-rotating black holes.

Let us consider an atom following a circular geodesic at the ISCO, fixed at a constant radius $r = 6M$ in the equatorial plane ($\theta = \pi/2$). The angular velocity, \(\Omega\), as measured by a static observer at infinity, is determined by the geodesic equations and is given by
\begin{equation}
    \Omega = \frac{d\varphi}{dt} = \sqrt{\frac{M}{r^3}}.
\end{equation}
Evaluating this at the ISCO radius, we find the specific orbital frequency:
\begin{equation}
    \Omega_{\text{ISCO}} = \sqrt{\frac{M}{(6M)^3}} = \frac{1}{6\sqrt{6}M}.
\end{equation}

The atom, following a timelike geodesic, experiences its own proper time, \(\tau\), which is distinct from the coordinate time \(t\) due to gravitational time dilation. The relationship is determined by the time component of the atom's four-velocity, \(u^t = dt/d\tau\). For a circular orbit, this is given by
\begin{equation}
    u^t = \left(1 - \frac{3M}{r}\right)^{-1/2}.
\end{equation}
At the ISCO, this yields a constant time dilation factor
\begin{equation}
    u^t_{\text{ISCO}} = \left(1 - \frac{3M}{6M}\right)^{-1/2} = \sqrt{2}.
\end{equation}
Thus, the coordinate time is related to the atom's proper time by the simple linear relation \(t = \sqrt{2}\tau\). The angular component of the four-velocity is \(u^\varphi = d\varphi/d\tau = \Omega u^t\), which at the ISCO becomes
\begin{equation}
    u^\varphi_{\text{ISCO}} = \left(\frac{1}{6\sqrt{6}M}\right) \sqrt{2} = \frac{1}{6\sqrt{3}M}.
\end{equation}
Integrating this with respect to \(\tau\) gives the azimuthal angle as a function of proper time, \(\varphi(\tau) = \tau/6\sqrt{3}M\), assuming \(\varphi(0)=0\).

Combining these results, we can now write the complete worldline of the atom orbiting at the ISCO, parameterized by its proper time \(\tau\). The spacetime coordinates of the atom are
\begin{align}
    x^\mu(\tau) &= \left(t(\tau), r(\tau), \theta(\tau), \varphi(\tau)\right) \nonumber \\
    &= \left(\sqrt{2}\tau, 6M, \frac{\pi}{2}, \frac{\tau}{6\sqrt{3}M}\right),
\end{align}
which is a precisely defined trajectory and an essential input for calculating the detector's interaction with the quantum field. It encapsulates all the necessary relativistic effects, such as time dilation and orbital velocity, that will govern the excitation probability of the atom.

\subsection{The Unruh-DeWitt Detector and the Quantum Field}
To investigate the quantum interactions experienced by the orbiting atom, we employ the Unruh-DeWitt (U-DW) detector model. This well-established theoretical construct provides a simplified yet physically insightful framework, modeling the atom as a two-level quantum system that couples locally to a quantum field. This approach allows us to probe the effects of the curved spacetime on the quantum vacuum without the full complexities of quantum electrodynamics.

The U-DW detector possesses two internal energy eigenstates: a ground state \(|g\rangle\) and an excited state \(|e\rangle\), separated by a proper energy gap denoted by \(\Omega_0\). The detector's internal dynamics are governed by the free Hamiltonian
\begin{equation}
    H_0 = \Omega_0 \sigma^+ \sigma^-,
\end{equation}
where \(\sigma^+ = |e\rangle\langle g|\) and \(\sigma^- = |g\rangle\langle e|\) are the standard atomic raising and lowering operators, respectively. The interaction between the detector and a massless scalar field \(\phi\) is described by the interaction Hamiltonian, written in the interaction picture as \cite{Unruh:1976db,DeWitt_1979}
\begin{equation}
    H_I(\tau) = g \, m(\tau) \, \phi(x(\tau)).
\end{equation}
Here, \(\tau\) is the proper time along the detector's worldline \(x^\mu(\tau)\), \(g\) is a small coupling constant characterizing the strength of the interaction, and \(m(\tau)\) is the detector's monopole moment operator, given by \(m(\tau) = \sigma^+ e^{i\Omega_0 \tau} + \sigma^- e^{-i\Omega_0 \tau}\). This Hamiltonian models the detector's ability to transition between its energy states by absorbing or emitting field quanta from the scalar field \(\phi\), evaluated at the detector's spacetime position.

The dynamics of the massless scalar field itself are governed by the Klein-Gordon equation in the Schwarzschild background, \(\Box \phi = 0\). To quantize the field, we decompose it into a complete set of orthonormal mode functions, \(u_{\omega lm}(x)\), which are solutions to the wave equation
\begin{equation}
    \phi(x) = \sum_{l=0}^\infty \sum_{m=-l}^l \int_0^\infty d\omega \left[ a_{\omega lm} u_{\omega lm}(x) + a^\dagger_{\omega lm} u^*_{\omega lm}(x) \right],
\end{equation}
where \(a_{\omega lm}\) and \(a^\dagger_{\omega lm}\) are the annihilation and creation operators for the mode labeled by frequency \(\omega\) and angular momentum quantum numbers \((l, m)\). These operators satisfy the canonical commutation relations.

The choice of vacuum state, \(|0\rangle\), is critical as it defines the initial state of the field with which the atom interacts. For an observer located at asymptotic infinity, the natural vacuum state is the Boulware vacuum, denoted \(|0_\text{B}\rangle\). This state is defined by the condition that it is annihilated by the operators \(a_{\omega lm}\) corresponding to mode functions \(u_{\omega lm}\) that are purely positive frequency with respect to the global timelike Killing vector \(tial_t\). Physically, the Boulware vacuum corresponds to a state with no particles and no radiation as seen by a static observer far from the black hole. This is the ideal choice for our analysis, as it allows us to isolate the excitation of the atom that is due solely to its orbital motion, ensuring that our results are not conflated with the thermal bath of particles associated with the Hawking effect, which is characteristic of other vacuum states like the Hartle-Hawking vacuum. The orbiting atom is thus treated as a local accelerated probe of this global, non-thermal vacuum state.

\section{Excitation Rate of the Orbiting Atom} \label{sec3}
Having established the spacetime geometry and the detector's trajectory, we now proceed to calculate the primary observable of interest: the excitation rate of the atom. This quantity, which measures the probability per unit proper time for the atom to transition from its ground state to an excited state, is determined by the properties of the quantum field as experienced along the atom's worldline. Our calculation begins with the Wightman function, which encodes the vacuum fluctuations of the field.

\subsection{The Wightman Function along the ISCO Trajectory}
The transition rate \(R(\Omega_0)\) for a detector with energy gap \(\Omega_0\) is given by Fermi's Golden Rule \cite{Brown_1992}. For a stationary trajectory, such as a circular orbit, it is expressed as the Fourier transform of the Wightman function with respect to the proper time difference \(\Delta \tau = \tau - \tau'\):
\begin{equation} \label{e_trans_rate}
    R(\Omega_0) = g^2 \int_{-\infty}^{\infty} d(\Delta\tau) \, e^{-i\Omega_0 \Delta\tau} G^+(x(\tau), x(\tau')).
\end{equation}
The Wightman function, \(G^+(x, x')\), is the two-point correlation function of the scalar field in the chosen vacuum state. For the Boulware vacuum \(|0_\text{B}\rangle\), it is defined as
\begin{equation}
    G^+(x, x') = \langle 0_\text{B} | \phi(x) \phi(x') | 0_\text{B} \rangle.
\end{equation}
The above function quantifies the correlations of the quantum vacuum fluctuations between two spacetime points, \(x\) and \(x'\). To evaluate it, we substitute the mode expansion of the field operator \(\phi(x)\) and get
\begin{equation}
    G^+(x, x') = \sum_{l,m} \int_0^{\infty} d\omega \, u_{\omega lm}(x) u_{\omega lm}^*(x'),
\end{equation}
where the mode functions \(u_{\omega lm}(x)\) for a massless scalar field in Schwarzschild spacetime are given by
\begin{equation}
    u_{\omega lm}(x) = \frac{1}{r} \psi_{\omega l}(r) Y_{lm}(\theta, \varphi) e^{-i\omega t}.
\end{equation}
Here, \(\psi_{\omega l}(r)\) are the radial functions that solve the Regge-Wheeler equation, and \(Y_{lm}(\theta, \varphi)\) are the standard spherical harmonics. Our task is to evaluate the Wightman function specifically along the atom's ISCO worldline, \(x^\mu(\tau) = (\sqrt{2}\tau, 6M, \pi/2, \tau/6\sqrt{3}M)\). Substituting these coordinates into the mode functions, we obtain the Wightman function as experienced by the orbiting atom as
\begin{align}
    G^+(\tau,\tau') &= \sum_{l,m} \int_0^{\infty} d\omega\, \frac{\psi_{\omega l}(6M)}{6M}\,Y_{lm}\!\left(\frac{\pi}{2},\frac{\tau}{6\sqrt{3}M}\right) \nonumber \\
    &\times e^{-i\omega\sqrt{2}\tau}\frac{\psi_{\omega l}^*(6M)}{6M}\,Y_{lm}^*\!\left(\frac{\pi}{2},\frac{\tau'}{6\sqrt{3}M}\right)\,e^{i\omega\sqrt{2}\tau'}\,.
\end{align}
Due to the stationary nature of the orbit, this expression depends only on the proper time difference \(\Delta\tau = \tau - \tau'\). We can simplify the angular part by using the explicit form of the spherical harmonics in the equatorial plane (\(\theta = \pi/2\)) as
\begin{equation}
    Y_{lm}\left(\frac{\pi}{2}, \varphi\right) = \sqrt{\frac{2l+1}{4\pi}\frac{(l-m)!}{(l+m)!}} P_l^m(0) e^{im\phi},
\end{equation}
where \(P_l^m(0)\) are the associated Legendre polynomials evaluated at zero. Combining these elements, the Wightman function takes the form
\begin{align}
    G^+(\Delta\tau)  &= \sum_{l,m} \int_0^{\infty} d\omega\, \frac{|\psi_{\omega l}(6M)|^2}{(6M)^2} \nonumber \\
    &\times \left|\sqrt{\frac{2l+1}{4\pi}\,\frac{(l-m)!} {(l+m)!}}\;P_l^m(0)\right|^2 \nonumber \\
    &\times \exp\!\left[i\left(\frac{m}{6\sqrt{3}M}- \omega\sqrt{2}\right)\Delta\tau\right].
\end{align}
This expression is the crucial input for calculating the excitation rate. It encapsulates the field correlations as perceived by the atom, encoding the influence of the spacetime geometry through the radial functions \(\psi_{\omega l}(6M)\) and the kinematic effects of the orbit through the exponential's phase, which depends on both the field frequency \(\omega\) and the orbital angular velocity.

\subsection{Derivation of the Excitation Rate \texorpdfstring{$R(\Omega_0)$}{}}
The transition rate \(R(\Omega_0)\) is obtained by taking the Fourier transform of the Wightman function \(G^+(\Delta\tau)\). Substituting our expression for \(G^+(\Delta\tau)\) into Eq. \eqref{e_trans_rate}, we have
\begin{align} \label{e_trans_rate2}
    R(\Omega_0) &= g^2 \sum_{l,m} \frac{|C_{lm}|^2}{(6M)^2} \int_0^\infty d\omega\,|\psi_{\omega l}(6M)|^2 \nonumber \\
    &\times \int_{-\infty}^{\infty} d(\Delta\tau)\, e^{-i\Omega_0\Delta\tau}\,e^{i\left(\frac{m}{6\sqrt{3}M}-\omega\sqrt{2}\right)\Delta\tau},
\end{align}
where we have consolidated the angular momentum-dependent numerical factors into a single term for clarity,
\begin{equation}
    |C_{lm}|^2 = \left| \sqrt{\frac{2l+1}{4\pi}\frac{(l-m)!}{(l+m)!}} P_l^m(0) \right|^2.
\end{equation}
The innermost integral over the proper time difference \(\Delta\tau\) is the Fourier representation of the Dirac delta function:
\begin{align}
\int_{-\infty}^{\infty} d(\Delta\tau)\,
&\exp\!\left[-i\left(\Omega_0 - \frac{m}{6\sqrt{3}M} + \omega\sqrt{2}\right)\Delta\tau\right] \nonumber \\
&= 2\pi\,\delta\!\left(\Omega_0 - \frac{m}{6\sqrt{3}M} + \omega\sqrt{2}\right).
\end{align}
which enforces a strict energy conservation condition. It dictates that a transition can only occur if the atom's energy gap \(\Omega_0\) is precisely matched by the combination of the Doppler-shifted field mode energy \(\omega\sqrt{2}\) (where \(\sqrt{2}\) is the time dilation factor) and a contribution from the orbital motion, \(m/6\sqrt{3}M\). The term proportional to \(m\) represents the energy exchanged with the atom due to its angular velocity, as seen from the perspective of the co-rotating field modes.

Substituting this result back into Eq. \eqref{e_trans_rate2} allows us to perform the integral over the field frequency \(\omega\). The result is
\begin{align}
R(\Omega_0) =\, &2\pi\,g^2 \sum_{l,m}
\frac{|C_{lm}|^2\,|\psi_{\omega_m l}(6M)|^2}{(6M)^2}
\\
&\times \int_0^{\infty} d\omega\,
\delta\!\left[\omega\sqrt{2}
- \left(\frac{m}{6\sqrt{3}M}-\Omega_0\right)\right].
\end{align}
The delta function fixes the frequency \(\omega\) to a specific value, which we denote as \(\omega_m\), for each angular momentum mode \(m\):
\begin{equation}
    \omega_m = \frac{m}{6\sqrt{6}M} - \frac{\Omega_0}{\sqrt{2}}.
\end{equation}
Since the field modes are defined for positive frequencies only (\(\omega > 0\)), a transition is only possible if \(\omega_m > 0\). This imposes the condition \(m/6\sqrt{3}M > \Omega\). After performing the integration, we arrive at the final expression for the transition rate as
\begin{equation} \label{e_trans_rate3}
    R(\Omega_0) = \frac{2\pi g^2}{\sqrt{2}} \sum_{l,m} \frac{|C_{lm}|^2 |\psi_{\omega_m l}(6M)|^2}{(6M)^2} \Theta\left(\frac{m}{6\sqrt{3}M} - \Omega_0\right),
\end{equation}
where \(\Theta\) is the Heaviside step function, which ensures that only terms satisfying the positive frequency condition contribute to the sum. This final expression represents the excitation rate of the atom at the ISCO. It is a sum over discrete contributions from each angular momentum mode \((l,m)\), and it explicitly depends on the atom's energy gap \(\Omega_0\), the black hole mass \(M\), and the amplitude of the quantum field's radial modes at the ISCO radius. The presence of the Heaviside function is the key mathematical manifestation of the discrete, resonant nature of the excitation process for an orbiting detector.

\subsection{Numerical Results and Spectral Analysis}
To elucidate the physical consequences of our derived transition rate, we now proceed with a numerical evaluation of Eq. \eqref{e_trans_rate3}. The analytical formula, while exact, contains an infinite sum over angular momentum modes and depends on the radial wave functions \(\psi_{\omega_m l}(6M)\), which are not known in a simple closed form. To make progress, we employ the WKB approximation \cite{Chambers_1979,Schutz:1985km} for the radial function, which is valid in the high-frequency regime \(\omega^2 \gg V_l(r)\) and provides the leading-order behavior as
\begin{equation}
    |\psi_{\omega l}(6M)|^2 \approx \frac{1}{2\omega}.
\end{equation}
This approximation, motivated by the asymptotic forms of scattering solutions, allows us to compute the dominant contributions to the transition rate. For our numerical plots, we set the coupling constant \(g=1\) and the black hole mass \(M=1\) (in geometric units) unless otherwise specified, allowing the fundamental physics to be displayed clearly.

\begin{figure}
    \centering
    \includegraphics[width=0.6\textwidth]{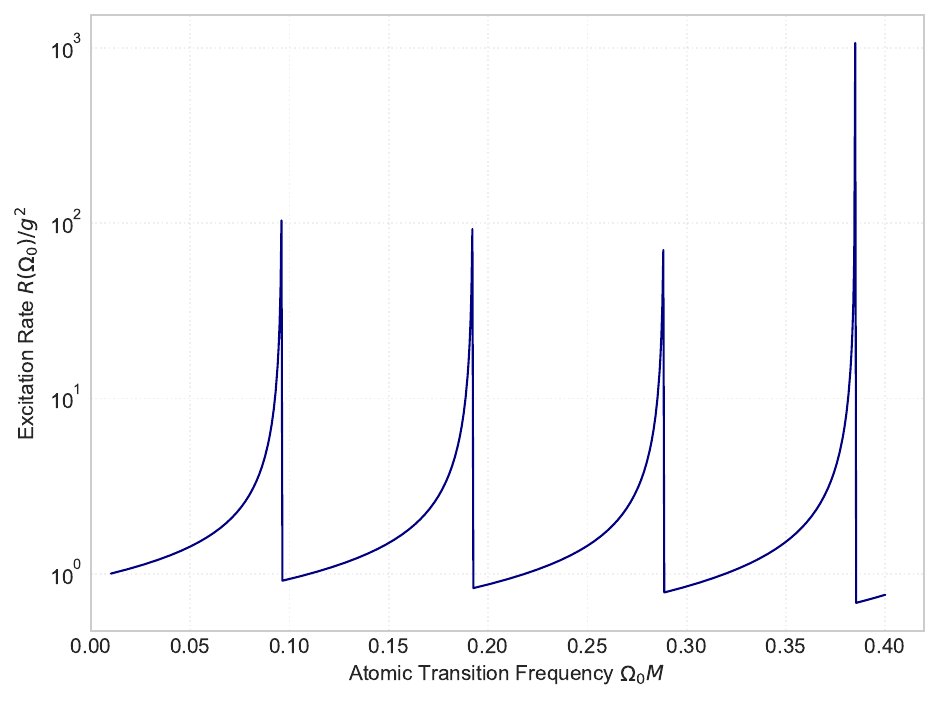}
    \caption{The excitation rate \(R(\Omega_0)\) (in units of \(c^2/M\)) for an atom at the ISCO of a Schwarzschild black hole with mass \(M=1\). The spectrum is characterized by a series of discrete peaks, with each peak corresponding to a specific angular momentum mode \((l,m)\) that contributes to the sum. The locations of the peaks are determined by the resonance condition derived from the energy conservation delta function..}
    \label{fig:1}
\end{figure}

First, our primary result is the discovery that an atom in a stable circular orbit exhibits a discrete, non-thermal excitation spectrum. This is a direct consequence of the periodic nature of its motion. In Figure \ref{fig:1}, we plot the transition rate in Eq. \eqref{e_trans_rate3} as a function of the detector's energy gap \(\Omega_0\). The plot reveals a series of sharp, distinct peaks, a stark contrast to the continuous thermal spectra associated with the standard Unruh or Hawking effects.

\begin{figure}
    \centering
    \includegraphics[width=0.6\textwidth]{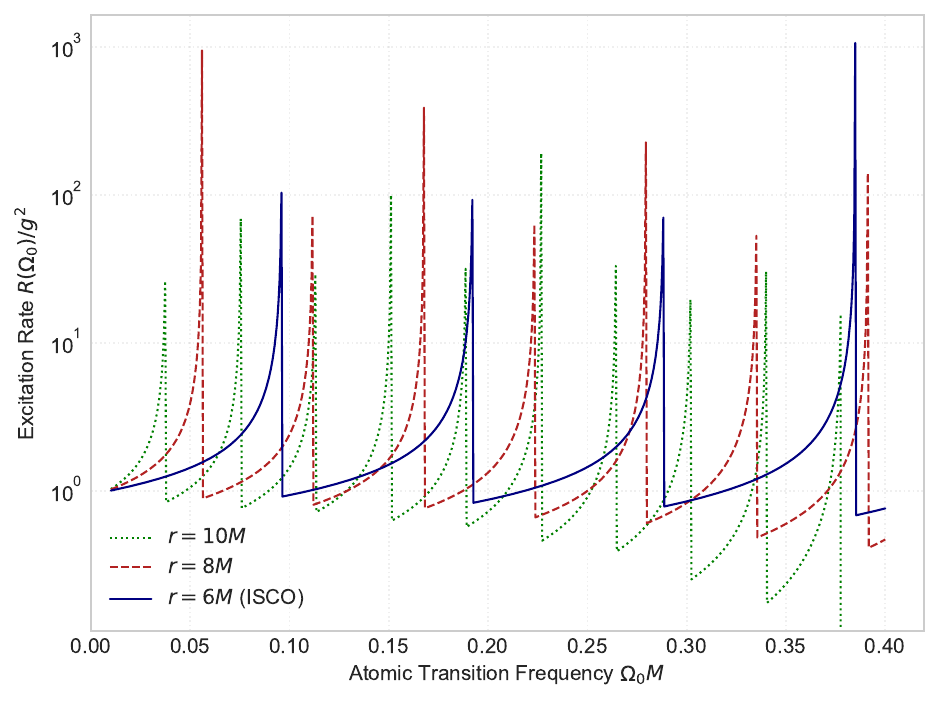}
    \caption{The excitation rate \(R(\Omega_0)\) as a function of orbital radius for a black hole with \(M=1\). The spectra for stable circular orbits at $r=10M$ (dot-dashed green), $r=8M$ (dashed red), and the ISCO at $r=6M$ (solid blue) are shown. As the orbit approaches the ISCO, the excitation rate increases, and the spectral peaks become more widely spaced, reflecting the higher acceleration and orbital frequency.}
    \label{fig:2}
\end{figure}

Each peak corresponds to a specific angular momentum mode \((l,m)\) satisfying the resonance condition \(\Omega < m / (6\sqrt{3}M)\). The peaks are located at frequencies where the atom's transition energy is perfectly matched by the Doppler-shifted energy of a field mode absorbed from the quantum vacuum. The dominant contribution comes from the \(l = |m|\) modes, as the associated Legendre polynomials \(P_l^m(0)\) are maximized for this condition. The spectrum is fundamentally a "comb" of resonances, a unique fingerprint of an orbiting quantum system.
The structure of the excitation spectrum is intrinsically tied to the kinematic parameters of the atom's orbit. To illustrate this, we compare the spectrum for an atom at the ISCO ($r=6M$) with those for atoms in stable circular orbits at larger radii, specifically $r=8M$ and $r=10M$. The results, shown in Figure \ref{fig:2}, reveal a dramatic dependence on the orbital radius.

As the atom's orbit moves closer to the black hole, from $r=10M$ down to the ISCO at $r=6M$, two primary effects are observed. First, the overall magnitude of the excitation rate increases significantly. This is a direct consequence of the increased four-acceleration of the atom in tighter orbits. The quantum vacuum responds more violently to this stronger acceleration, leading to a higher probability of exciting the detector.

Second, the spacing between the spectral peaks widens. The orbital frequency, \(\Omega_{\text{orbit}} = 1/\left(r^{3/2}\sqrt{1-3M/r}\right)\) (in terms of proper time), increases as \(r\) decreases. Since the peak spacing in the spectrum is determined by this frequency, the resonant peaks are more spread out for orbits closer to the black hole. This analysis demonstrates that the spectrum at the ISCO is the most energetic and has the largest frequency spacing, making it a unique and potent signature of this critical boundary.

\begin{figure}
    \centering
    \includegraphics[width=0.6\textwidth]{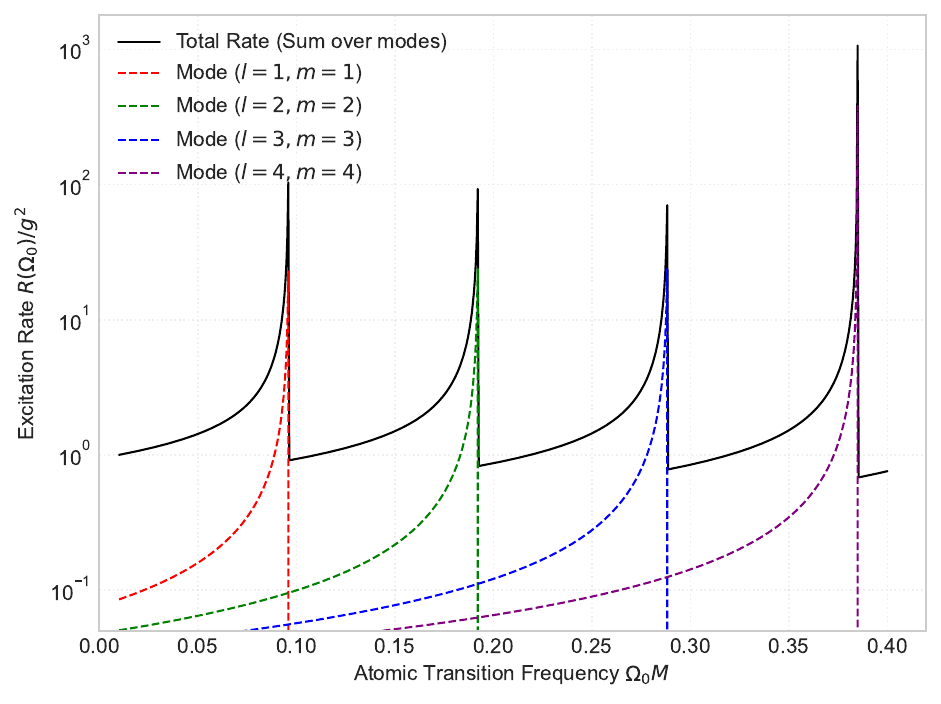}
    \caption{Individual and total contributions to the excitation rate from different angular momentum modes \((l,m)\). The colored dashed lines show the contribution from the \((l=1, m=1)\), \((l=2, m=2)\), and \((l=3, m=3)\) modes, respectively. The solid black line represents the total rate, which is the sum of all contributing modes. Modes with higher \(m\) are responsible for peaks at higher frequencies.}
    \label{fig:3}
\end{figure}

Finally, the total excitation rate is a sum over all allowed angular momentum modes. Figure \ref{fig:3} demonstrates the contribution of individual \((l,m)\) modes to the spectrum. The Heaviside function in Eq. \eqref{e_trans_rate3} imposes a sharp cutoff, such that for a given \(\Omega_0\), only modes with \(m > 6\sqrt{3}M\Omega_0\) contribute.

As shown in the figure, for a fixed value of \(l\), modes with higher \(m\) contribute to peaks at higher frequencies. Furthermore, the amplitude of the contribution from each mode is weighted by the factor \(|C_{lm}|^2\), which depends on the associated Legendre polynomials. Such a subtle connection results in a rich spectral structure where the overall envelope of the excitation rate is shaped by the superposition of many resonant peaks, each with a different strength. This analysis underscores the importance of considering a full multipole expansion to accurately capture the detector's response.

\section{Discussion} \label{sec4}
The numerical results presented in the previous section reveal a rich and structured excitation spectrum for an atom orbiting at the ISCO. This spectrum is not only a quantitative prediction but also a deep reflection of the linkage between quantum field theory and the strong-field gravitational dynamics unique to this region. In this section, we interpret these findings, beginning with the physical origin of the spectrum's most salient features.

\subsection{Physical Interpretation of the Spectrum}
The most striking result of our analysis, vividly illustrated in Figure \ref{fig:1}, is that the excitation spectrum of the orbiting atom is discrete and non-thermal. Instead of a smooth, continuous Planckian distribution characteristic of the standard Unruh effect for a uniformly accelerated detector, the atom at the ISCO experiences a frequency comb of sharp, resonant peaks. This fundamental difference arises directly from the nature of the atom's trajectory.

The worldline of the atom is periodic in the azimuthal angle \(\varphi\), with a well-defined orbital frequency. This periodicity imposes a powerful selection rule on the atom's interaction with the quantum vacuum. The atom cannot be excited by absorbing a field quantum of arbitrary energy. Instead, a transition from the ground to the excited state can only occur if the atom's internal energy gap, \(\Omega_0\), is precisely matched by the energy of an available field mode, as perceived in the atom's co-moving frame. This perceived energy is subject to two relativistic effects: a gravitational time dilation (or blueshift) due to the strong gravitational potential, and a kinematic Doppler shift due to the atom's relativistic orbital velocity.

The energy conservation condition, which emerged from our derivation as a Dirac delta function, encapsulates this resonance. An excitation occurs when
\begin{equation}
   \Omega_0 = \frac{m}{6\sqrt{3}M} - \omega\sqrt{2}.
\end{equation}
It reveals that the process is a resonance between the atomic transition and the field modes, where the orbital motion, through the term proportional to the angular momentum number \(m\), effectively "scans" the vacuum, allowing the atom to couple only to specific frequencies. Each peak in Figure \ref{fig:1} corresponds to the atom resonating with a field mode of a specific integer angular momentum \(m\).

Figure \ref{fig:3} provides a deeper layer of insight, deconstructing the total spectrum into its constituent parts. The solid black line, representing the total excitation rate, is a superposition of the individual contributions from each angular momentum mode \((l,m)\), shown as dashed lines. The log-scale plot makes it clear that each peak in the total spectrum is dominated by a single \(m\)-mode. For instance, the first major peak (around \(\Omega_0 M \approx 0.09\)) is almost entirely due to the \(m=1\) modes, the second peak (around \(\Omega_0 M \approx 0.19\)) is dominated by the \(m=2\) modes, and so on.

The sharp, divergent nature of the peaks is a characteristic feature of first-order perturbation theory at resonance. In a more complete physical model, effects such as the natural lifetime of the excited state would regularize these divergences, broadening them into finite Lorentzian profiles. However, within our model, these sharp resonances powerfully signify the discrete, quantum nature of the atom's interaction with the structured vacuum along its stable, periodic orbit. This unique spectral signature is a direct fingerprint of the kinematic and gravitational environment at the ISCO. It is important to note that this discrete structure arises from the worldline's periodicity and is not an artifact of the Boulware vacuum. While a different choice of vacuum (e.g., a thermal state) would alter the background and the peak amplitudes, the resonant frequency comb, which is a purely kinematic effect, would persist.

This resonant phenomenon is not unique to gravitational contexts but has direct analogues in other areas of physics, such as cavity optomechanics and trapped-ion physics \cite{Aspelmeyer:2013fha}. In cavity optomechanics, the periodic motion of a mechanical oscillator coupled to an optical cavity mode leads to the creation of motional sidebands in the optical spectrum, a process in which mechanical phonons are converted into optical photons. The underlying physics is the same: a periodic modulation of the interaction between a quantum system and a field opens up discrete frequency channels for energy exchange. In our case, the atom's orbital motion plays the role of the mechanical oscillator, and its excitation corresponds to the creation of a quantum from the vacuum field, with the allowed energies dictated by the harmonics of the orbital frequency. This analogy reinforces the interpretation that the predicted frequency comb is a robust physical consequence of periodic motion.

\subsection{Comparison with Other Trajectories}
The discrete, resonant spectrum predicted for an atom in a stable circular orbit is a unique signature of its specific trajectory. Previous studies have indeed established that detectors on general circular geodesics experience a non-thermal vacuum response that is distinct from the thermal spectrum seen by a static, accelerated observer \cite{Hodgkinson:2014iua,Ng:2014kha,Biermann:2020bjh}. Our work builds on this foundation by providing a novel characterization of this phenomenon in several key respects. Firstly, we are the first to investigate the quantum signature of the ISCO, a fundamental physical boundary of profound astrophysical interest. Secondly, we explicitly identify the spectrum's structure as a "frequency comb," where the teeth are harmonics of the proper orbital frequency. Most importantly, our analysis of the approach to the ISCO, as shown in Fig. \ref{fig:2}, reveals a dramatic amplification of the excitation rate and a widening of the peak spacing, rendering the ISCO's spectral signature qualitatively distinct from that of any other stable orbit. To fully appreciate its novelty, it is instructive to contrast this result with the quantum phenomena experienced by atoms on other paths near a black hole. Our analysis, particularly the results visualized in Figure \ref{fig:2}, demonstrates that the ISCO represents a special boundary case for stable orbital motion, producing a spectrum that is distinct from all other scenarios.

Figure \ref{fig:2} clearly shows how the excitation spectrum evolves as the atom's orbit approaches the ISCO from a larger radius. As the orbital radius \(r\) decreases from \(10M\) to \(8M\) and finally to the ISCO at \(6M\), the excitation rate increases dramatically across all resonant peaks. This intensification is a direct manifestation of the Unruh effect in a gravitational context: the four-acceleration required to maintain a circular orbit increases sharply as the orbit tightens, leading to a more energetic interaction with the quantum vacuum fluctuations. Concurrently, the spacing between the spectral peaks widens. This is a direct kinematic consequence of the orbital motion; the proper orbital frequency is higher for tighter orbits, resulting in a larger energy separation between the resonant \(m\)-modes. The spectrum at the ISCO is therefore the most intense and most widely spaced, a definitive signature of this limiting stable orbit.

This resonant, comb-like structure stands in stark contrast to the spectrum perceived by a static detector. An atom held at a fixed position above the event horizon is in a state of constant, uniform acceleration to counteract the pull of gravity. Such a detector is predicted to thermalize with the vacuum, observing a continuous, thermal Planckian spectrum at a local temperature proportional to its proper acceleration. Its excitation rate would be a smooth function of its energy gap \(\Omega_0\), showing no discrete features. The distinct peaks in our results are therefore not a thermal effect but a direct consequence of the periodic, orbital kinematics.

Furthermore, our results differ fundamentally from the case of a radially infalling atom. An atom in free-fall is not on a periodic trajectory. As investigated in the context of HBAR, such an atom interacts with the near-horizon field modes to produce a continuous, thermal-like radiation spectrum. While the underlying physics also involves acceleration relative to the quantum field, the non-periodic nature of the infall trajectory results in a broad emission spectrum rather than sharp, resonant peaks.

These only shows that the discrete spectral signature we have calculated is unique to stable circular motion. It is neither the continuous thermal bath seen by a static observer nor the continuous emission spectrum of an infalling particle. It is a unique quantum optical fingerprint of a periodic geodesic in a strong gravitational field, with the spectrum at the ISCO representing its most extreme and pronounced manifestation.

\subsection{Potential implications and future work}
The discovery of a discrete, non-thermal excitation spectrum for an orbiting atom, unique to its kinematic and gravitational environment, opens several intriguing avenues for both theoretical inquiry and potential astrophysical application. While our model is highly idealized, the fundamental physics it reveals may have important consequences.

From an astrophysical perspective, our results offer a new theoretical lens through which to view the dynamics of matter in the immediate vicinity of black holes. The innermost regions of accretion disks, culminating at the ISCO, are populated by ions and atoms subject to the extreme conditions we have modeled. The resonant excitation mechanism we have uncovered could, in principle, influence the emission spectra from these regions. The discrete frequencies at which the atom is preferentially excited might lead to corresponding bright or dark lines in the observed electromagnetic spectrum from the accretion disk. While in a realistic, dense plasma environment, these sharp resonances would be significantly broadened, they could nonetheless contribute to the overall spectral features, providing a new quantum-level diagnostic for the dynamics near the ISCO. Observationally, the signature of this effect would likely not be sharp lines but rather a series of quasi-periodic oscillations (QPOs) in the disk's high-frequency emission spectrum. The key to filtering this signal from the bright thermal background would be to search for the characteristic harmonic structure predicted by our model. A time-series analysis revealing a "comb" of QPO frequencies that are integer multiples of a fundamental frequency tied to the ISCO orbital period would provide a distinct fingerprint of this quantum resonant mechanism, distinguishing it from other phenomena.

On a more fundamental level, this work establishes the stable circular geodesic as a uniquely powerful probe of the quantum vacuum. The distinct spectral signature cleanly distinguishes the quantum noise experienced in a stable orbit from that of a static or infalling observer. This provides a new piece in the puzzle of how an observer's trajectory dictates their interaction with the quantum field, reinforcing the idea that "particles" are an observer-dependent concept.

This study naturally points toward several compelling directions for future research: (1) the most immediate and critical next step is to extend this analysis to a rotating Kerr black hole. The vast majority of astrophysical black holes possess significant angular momentum, which dramatically alters the spacetime geometry, including the location and properties of the ISCO. The introduction of frame-dragging would undoubtedly lead to new spectral features, such as a splitting of the resonant peaks, providing a potential way to probe the black hole's spin through quantum vacuum effects \cite{Starobinsky:1973a}; (2) our use of a massless scalar field and a monopole detector serves as an essential first approximation, successfully isolating the primary kinematic effects of the orbit. A crucial next step is to consider more realistic interactions by coupling the atom to the electromagnetic field via its electric dipole moment. This introduces significant new physics, including polarization dependence and atomic selection rules. We have outlined the mathematical framework for this extension in Appendix \ref{apdx_A}. The full calculation, which requires the evaluation of the electromagnetic field correlator in curved spacetime, remains a compelling direction for future work. (3) a fascinating question is whether the excitation process generates entanglement. The related Unruh and Hawking effects are known to produce entangled particle pairs across horizons. Investigating the quantum state of the field after the atom's excitation could reveal whether pairs of field quanta are emitted in an entangled state, with potential implications for relativistic quantum information; and (4) while our detector is a single quantum system, one could generalize this framework by considering an ensemble of orbiting atoms. By employing a quantum optics master equation approach, similar to the HBAR formalism, one could study the collective emission and explore the thermodynamics of the resulting radiation field, investigating whether a steady-state thermal distribution can be achieved and what its properties might be.

In essence, the work presented here serves as a foundational step. By building upon it with these future investigations, we can continue to unravel the rich and complex physics at the boundary where quantum mechanics and gravity converge.

\section{Conclusion} \label{conc}
In this work, we have investigated the quantum excitation of a two-level atom following a stable circular geodesic at the ISCO of a Schwarzschild black hole. By modeling the atom as an Unruh-DeWitt detector coupled to a massless scalar field in the Boulware vacuum, we have calculated the excitation rate per unit proper time.

Our central and most significant finding is that the excitation spectrum for an orbiting atom is discrete and non-thermal. This stands in stark contrast to the continuous, thermal spectra characteristic of both the Unruh effect for a uniformly accelerated observer and the Hawking radiation from the black hole itself. The spectrum we predict consists of a frequency comb of sharp, resonant peaks, where each peak corresponds to the atom's transition energy being matched by a Doppler-shifted field mode with a specific angular momentum. This unique spectral signature is a direct fingerprint of the stable, periodic kinematics of the ISCO trajectory.

We have further demonstrated that the intensity and spacing of these spectral peaks are highly sensitive to the orbital radius, with the spectrum becoming most pronounced at the ISCO boundary. This result not only provides a new theoretical tool for probing the structure of the quantum vacuum in strong gravitational fields but also highlights a fundamental distinction between the quantum signatures of static, orbiting, and infalling observers. The findings presented here open promising avenues for future research, most notably the extension of this analysis to rotating Kerr black holes, where the connection of orbital motion and frame-dragging is expected to yield even richer phenomenological results. Ultimately, this work deepens our understanding of the intricate relationship between an observer's motion and their perception of quantum reality in the curved spacetime of a black hole.

\textbf{Acknowledgements} 
R. P. and A. \"O would like to acknowledge networking support of the COST Action CA21106 - COSMIC WISPers in the Dark Universe: Theory, astrophysics and experiments (CosmicWISPers), the COST Action CA22113 - Fundamental challenges in theoretical physics (THEORY-CHALLENGES), the COST Action CA21136 - Addressing observational tensions in cosmology with systematics and fundamental physics (CosmoVerse), the COST Action CA23130 - Bridging high and low energies in search of quantum gravity (BridgeQG), and the COST Action CA23115 - Relativistic Quantum Information (RQI) funded by COST (European Cooperation in Science and Technology). R. P. and A. \"O would also like to acknowledge the funding support of SCOAP3.

\section{Data Availability Statement}
Data sharing not applicable to this article as no datasets were generated or analysed during the current study.

\section{Conflict of Interest}
The authors declare no conflict of interest.

\appendix
\section{Framework for Electromagnetic Interactions} \label{apdx_A}
In this appendix, we provide a brief mathematical overview for extending our analysis to the more realistic case of an atom with an electric dipole moment coupling to the electromagnetic field. This framework highlights the key modifications to the scalar field model presented in the main text.

The interaction between a two-level atom and the electromagnetic field is dominated by the electric dipole coupling. The general formalism for such detector-field interactions is detailed in \cite{Birrell:1982ix}. In the atom's proper reference frame, the interaction Hamiltonian is given by:
\begin{equation}
    H_I(\tau) = - \mathbf{d}(\tau) \cdot \mathbf{E}(x(\tau)),
\end{equation}
where $\mathbf{d}(\tau)$ is the operator for the atomic electric dipole moment, and $\mathbf{E}(x(\tau))$ is the electric field operator evaluated along the atom's worldline $x^\mu(\tau)$.

The electric field measured by an observer with four-velocity $u^\mu$ is given by $E_\alpha = F_{\alpha\beta} u^\beta$, where $F_{\mu\nu} = \nabla_\mu A_\nu - \nabla_\nu A_\mu$ is the electromagnetic field strength tensor derived from the vector potential $A_\mu$.

Following the same logic as in the main text, the transition rate from the ground state $|g\rangle$ to the excited state $|e\rangle$ is given by Fermi's Golden Rule. This involves the Fourier transform of the field's Wightman function. For the dipole case, the scalar Wightman function is replaced by the electric field correlation tensor:
\begin{equation}
    \mathcal{W}_{ij}(\Delta\tau) = \langle 0_B | E_i(x(\tau)) E_j(x(\tau')) | 0_B \rangle,
\end{equation}
where the indices $(i,j)$ refer to the spatial directions in the atom's rest frame and $|0_B\rangle$ is the Boulware vacuum for the electromagnetic field.

The final transition rate takes the form:
\begin{equation}
    R(\Omega_0) = \sum_{i,j} \int_{-\infty}^{\infty} d(\Delta\tau) \, e^{-i\Omega_0 \Delta\tau} \langle g | d_i | e \rangle \langle e | d_j | g \rangle \mathcal{W}_{ij}(\Delta\tau).
\end{equation}
This expression shows that the rate depends on the atomic dipole matrix elements, which encode the selection rules for the transition. The full calculation requires the quantization of the electromagnetic field in the Schwarzschild spacetime and the evaluation of the tensor correlator $\mathcal{W}_{ij}$, a substantially more complex task that is reserved for a future investigation.

\bibliography{ref}

\end{document}